# Spin Wave Magnetic NanoFabric: A New Approach to Spin-based Logic Circuitry

Alexander Khitun, Mingqiang Bao, and Kang L. Wang

Device Research Laboratory, Electrical Engineering Department,

FCRP Focus Center on Functional Engineered Nano Architectonics (FENA),

Nanoelectronics Research Initiative - The Western Institute of Nanoelectronics (WIN).

**Abstract**

We propose and describe a magnetic NanoFabric which provides a route to building reconfigurable spin-based logic circuits compatible with conventional electron-based devices. A distinctive feature of the proposed NanoFabric is that a bit of information is encoded into the phase of the spin wave signal. It makes possible to transmit information without the use of electric current and utilize wave interference for useful logic functionality. The basic elements include voltage-to-spin wave and wave-to-voltage converters, spin waveguides, a modulator, and a magnetoelectric cell. As an example of a magnetoelectric cell, we consider a two-phase piezoelectric-piezomagnetic system, where the spin wave signal modulation is due to the stress-induced anisotropy caused by the applied electric field. The performance of the basic elements is illustrated by experimental data and results of numerical modeling. The combination of the basic elements let us construct magnetic circuits for NOT and Majority logic gates. Logic gates AND, OR, NAND and NOR are shown to be constructed as the combination of NOT and a reconfigurable Majority gates. The examples of computational architectures such as Cellular Automata, Cellular Nonlinear Network and Field Programmable Gate Array are described. The main advantage of the proposed NanoFabric is in the ability to realize logic gates with less number of devices than it required for CMOS-based circuits. Potentially, the area of the elementary reconfigurable Majority gate can be scaled down to $0.1\mu m^2$. The disadvantages and limitations of the proposed NanoFabric are discussed.

I. Introduction

The rapid approach to the scaling limit of metal-oxide semiconductor field-effect transistor (MOSFET) stimulates a great deal of interest to new technologies able to overcome the constrains inherent to CMOS-based circuitry and provide a route to more functional and less power dissipating logic devices. As one of the possible directions, Spintronics has been recognized as a new emerging approach to electronics in the semiconductor community [1]. The utilization of spin opened new horizons for semiconductor device development which resulted in a number of promising ideas [2-5] All proposed spin-based devices can be classified in two categories. The first category includes devices such as Magnetoresistive Random Access Memory (MRAM) [6], Magnetic Tunneling Junction (MTJ) [7], Spin-Modulator proposed Datta and Das [2], where spin is used to control electric current. The second one comprises of spin-based devices such as Magnetic Cellular Automata [8], Domain-Wall Logic [9], and devices with Spin Wave Bus [10] which operation does not relay on electric current. Both these directions have advantages and limitations associated with scalability, power dissipation, defect tolerance, and compatibility with CMOS-based circuits. However, the replacement of electric current by other physical mechanisms such as dipole-dipole or exchange interaction may be more promising as it does not possess constrains inherent to CMOS-based technology and offers new approaches for logic circuit construction. Today, the area of magnetic memory is much more elaborated in comparison to the magnetic-based logic. One of the major problems of the spin-based logic circuitry is device variability. For example, the variation of size and position of magnetic domains in MCA results in error probability increase and leads to power dissipation [11]. That would be of great practical importance to develop reconfigurable spin-based logic elements for integration with spin-based memory, and develop a magnetic *NanoFabric* as a universal tool for building defect-tolerant spin-based logic circuits.

NanoFabric is a special design methodology aimed to resolve the variability issue by using reconfigurable elements [12]. A nanofabric may be a two-dimensional mesh of interconnected reconfigurable logic blocks, which can be used to route the information flow. The ability to reconfigure basic blocks is the key requirement, which makes NanoFabric resistant to defects and make use of the maximum system resources. Using NanoFabrics, there is a large dependence based on the architecture to overcome the constraints associated with direct assembly of particular nanometer-scale components [13]. For example, different types of NanoFabrics have been developed for molecular-based devices [14], nanotube arrays [15], and domain wall logic [9]. In this work we present a magnetic NanoFabric, which consists of spin-based devices and utilizes spin waves for information transmission and processing.

In the preceding works [10, 16, 17], we have developed the general concept of logic circuits with Spin Wave Bus. Briefly, the basic idea to use magnetic films as spin conduit of wave propagation or referred to as – Spin Wave Bus, where the information can be coded into a phase of the propagating spin wave. There are several advantages of this approach: (i) information transmission is accomplished without an electron transport; (ii) ability to use superposition of spin waves in the bus to achieve useful logic functionality;

(iii) a number of spin waves with different frequencies can be simultaneously transmitted among a number of spin-based devices; (iv) the interaction between spin waves and the outer devices can be done in a wireless manner, via a magnetic field. The examples of logic functionality were demonstrated by numerical modeling [10]. A first working spin-wave based logic circuit has been experimentally demonstrated by M. Kostylev et al. [18]. In all these initial works, the most of the attention was paid to the feasibility study of spin wave utilization for analog computing. It is needless to say, that the ability to provide computation in a digital form is an essential requirement for modern computational architectures. In this work, for the first time, we propose and describe a combination of spin wave bus with a magnetoelectric material structure to convert spin wave signal in a digital form.

Currently, there is a growing interest to multiferroic and magnetoelectric materials primary for memory application [19]. Multiferroics is a special type of materials that possesses simultaneously electric and magnetic orders [20, 21], which provides the unique opportunity to control its magnetic properties by applied electric field. However, there are only a few room temperature multiferroic materials known today [21], e.g. $BiFeO_3$ and its derivatives. An alternative method for obtaining artificial structure with magnetoelectric effect is to couple between two materials such as a ferromagnetic and a ferroelectric [22]. In this case, the magnetoelectric coupling arises as a combined effect of two: piezoelectricity and piezomagnetism. An example of a voltage control of magnetization easy-axes in a magnetoelectric system consists of ferromagnetic/piezoelectric films has been recently experimentally demonstrated [23].

The use of magnetoelectric structure in combination with spin wave bus has many potential advantages, some of them we describe in the present work. The rest of the paper is organized as follows. In the next section we describe basic devices of the proposed magnetic NanoFabric and logic circuits based on these devices. In Section III we present experimental data and results of numerical simulation illustrating the performance of these devices. Section IV is devoted to possible architectures that can be constructed by the tools of the NanoFabric. Discussions and Conclusions are given in sections V and VI, respectively.

II. Basic elements and logic circuits

The operation of the spin wave-based logic circuit includes three major steps: (i) conversion of the input data into spin wave signals, (ii) computation using spin waves, and (iii) conversion the result of computation in the form of voltage signals. The distinctive feature of the spin wave-based logic circuit is that a bit of information is encoded into the *phase* of the spin wave. Two relative phases of "0" and "$\pi$" may be used to represent two logic states 1 and 0. Being excited, spin waves propagate throughout the structure consists of spin waveguides. The data processing in the circuit is accomplished by manipulating the relative phases of the propagating spin waves. Finally, the result of computation is converted in the voltage signal. There are four basic elements in the proposed NanoFabric: converters, spin waveguides, modulators, and magnetoelectric cells. In Fig.1 we schematically show the material structure for each element and depict a symbol for circuit notification.

*Converter* is aimed to translate input voltage pulse into spin wave signal and vice versa. The simplest example of the converter is a conducting loop placed on the top of the magnetic film (as shown in Fig.1(a)). The loop and the film are separated by an insulating layer. The applied voltage produces electric current trough the loop. In turn, the current generates magnetic field, which excites spin waves in the film. The polarity of the input voltage pulse defines the direction of the current through the loop (clockwise or contrawise), and, thus, defines the initial phase (0 or $\pi$) of the excited spin wave signal. A propagating spin wave changes the local magnetization of the magnetic material and alters the magnetic flux. According to the Faraday's law, the change of the magnetic flux through the surface generates an inductive voltage proportional to the rate of the magnetic flux change $E_{ind} = -d\Phi_m/dt$. The same conducting loop can be used for the detection of the inductive voltage produced by spin wave. Coplanar waveguides microstructures are widely used for spin wave excitation/detection and the detailed structure description can be found everywhere [24-26].

*Spin waveguides* are for the transmission of spin waves through the circuit. Its utility is similar to optical waveguides aimed to guide the propagation of electromagnetic waves. The waveguide structure may consist of a magnetic film, a wire or a combination of wires made of ferromagnetic, antiferromagnetic, or ferrite material. Spin wave signal can be divided or combined by the waveguide structures schematically shown in Fig.1(b). For example, an yttrium iron garnet (YIG) spin wave splitters and combiners were used in the first prototype logic circuit [18]. The particular choice of a magnetic material is defined by many physical and technological conditions including compatibility with silicon technology. Without loss of generality, hereafter, we will use ferromagnetic as a waveguide material.

*Modulator* (or spin wave phase modulator) is a device designed to provide $\pi$-phase shift to the propagating spin wave as a function of the applied voltage. As the spin wave transport is sensitive to external magnetic field, it is possible to use for modulation a local magnetic field produced by an electric current in a single conductor. In Fig.1(c) we have schematically shown a wire on the top of the waveguide structure placed perpendicularly to the direction of spin wave propagation. The operation of the modulator is controlled by the applied voltage $V_C$. We consider two states of modulator operation. In the "Off" state, there is no control voltage applied ($V_C=0$), there is no current through the wire, and there is no effect on spin wave propagation in the waveguide. In the "On" state, there is an applied voltage ($V_C=V_\pi$), so the modulator provide $\pi$-phase shift to the propagating spin wave.

*Magnetoelectric cell* is a device to control spin wave signal by the applied electric field via the magneto-electric coupling. There is a special type of materials, so called multiferroics, that possesses simultaneously electric and magnetic orders and show prominent magneto-electric couplin [20, 21]. However, there are only a few room temperature multiferroic materials known today [21], e.g. $BiFeO_3$ and its derivatives. An alternative method for engineering a magnetoelectric structure is to use a two-phase system comprising two materials such as a piezoelectric and a piezomagnetic [22]. Each material may be independently optimized for room temperature performance. There are several promising candidates for the piezoelectric-piezomagnetic structure, which have been experimentally studied showing prominent magnetoelectric coupling: PZT/NiFe$_2$O$_4$ (1,400 mV cm$^{-1}$ Oe$^{-1}$)[27], CoFe$_2$O$_4$/BaTiO$_3$ (50 mV cm$^{-1}$ Oe$^{-1}$)[28], PZT/Terfenol-D (4,800 mV cm$^{-1}$ Oe$^{-1}$)[29]. The bias voltage can be applied across the structure using the gate electrode and a conducting ferromagnetic layer (CoFe, or NiFe, for example). The stress produced by the piezoelectric results in the easy axis rotation in the piezomagnetic material [30]. In Fig.1(d) we have schematically shown a magnetoelectric cell consists of a piezoelectric (e.g. PZT) and a piezomagnetic/ferromagnetic (e.g. CoFe). The operation of the cell is controlled by the applied voltage $V_H$. In the "Off" state ($V_H=0$), the piezoelectric material has no effect on spin wave propagation in the ferromagnetic waveguide. In the "On" state ($V_H \neq 0$), the voltage applied to the piezoelectric material (e.g. PZT) produces mechanical stress, which changes the easy axis direction. We assume that the easy axis of the ferromagnetic material to be rotated on 90 degree at some control voltage $V_H=V_A$. In this state, the domains in the ferromagnetic material have two preferable directions of magnetization: along with or opposite to the "new" easy axis. As we show in the next section by numerical modeling, the particular direction can be defined by the phase of the spin wave propagating in the ferromagnetic film.

It is important to note that the magnetoelectric cell has several useful functions. First, it can work as a memory cell in the "On" state as long as the control voltage is applied. Second, it can be used as a spin wave amplifier. In the next section we present the result of numerical simulations showing spin wave amplitude increase as a result of the easy axis rotation. Third, the cell can be also used as a converter and as a spin wave modulator. In both cases, an effective magnetic field is produced not by an electric current but via the magnetoelectric effect. It is experimentally demonstrated an electrically and magnetically tunable microwave planar resonator consisting of yttrium iron garnet (YIG) and ferroelectric barium strontium titanate (BST) thin films [31], where the magnetic tuning had been achieved by the electric field. An artificial two-phase magnetoelectric CoPd/PZT cell was experimentally demonstrated by K. Sang-Koog et al, and the easy axis rotation up to 150 degree was observed [23]. These experimental data show the plausibility of using magnetoelectric cells in high frequency range (GHz) and at room temperature.

Next, we describe logic circuits that can be built by the basic elements. In Fig.2 we show logic gates and corresponding magnetic circuits. The use of spin wave phase for information encoding implies a specific approach to logic function realization. For example, a ferromagnetic wire of length $l=\lambda \cdot n$, ($n=1,2,3,..$ and $\lambda$ - is the spin wavelength) represents a Buffer gate. By analogy, a ferromagnetic wire of length $l=1/2 \cdot \lambda \cdot n$, ($n=1,2,3,..$) can be used as an Inverter. Of course, this simplified consideration does not take into account signal attenuation during signal propagation in the ferromagnetic waveguide. In order to compensate amplitude damping, an amplifier (the magnetoelectric cell) should be added. A multifunctional Buffer/Inverter one-bit logic gate can be constructed by combining input Converter, Modulator, Magnetoelectric Cell, and output Converter as it is shown in Fig.2(a). The length of the whole circuit is assumed to be an integer number of the spin wavelength. The circuit works as a Buffer gate if the Modulator is in the "Off" state, and it works as an Inverter if the Modulator is in the "On" state. The Converters are needed to provide the compatibility with the outer electron-based devices. Hereafter, we depict input and output Converters for each circuit, though no converters are needed if the circuits are integrated within the NanoFabric.

The utilization of wave interference provides an effective way to build a Majority logic gate. The interference of several spin waves results in the magnetization change which is the sum of the separate amplitudes of each wave. Similar to other waves, spin waves can interfere in a constructive (in phase) or a destructive (out of phase) manner. The magnetization change as a result of two spin waves interference has been considered in [10] and is not reproduced here. The amplitude of the resultant signal depends on how many waves are coming in phase or out of phase. In order to equalize the output amplitude, the magnetoelectric cell should be included. In Fig.2(b) we depicted a three-input Majority gate consists of Converters, Combiner, and the Magnetoelectric Cell. Three input spin wave signals of the same frequency are generated by the input Converters. The signals are mixed by the Combiner, and the output signal is amplified and equalized by the Magnetoelectric Cell. There is no restriction on the number of input nodes for the Majority gate. Moreover, a number of input Converters can be placed on the same waveguide line to minimize the circuit area.

A multifunctional logic gate can be built on the base of the described Majority gate by introducing two Modulators and by reserving one input for the control signal as it is shown in Fig.2(c). In this case, the whole circuit can be considered as a two-terminal reconfigurable gate. The control input is always generate spin wave with 0 initial phase (logic state 1). Depends on the control voltages applied to the Modulators, the gate operates as AND gate ($V_{C1}=V_{C2}=0$); OR gate ($V_{C1}=V_\pi, V_{C2}=0$); NAND gate ($V_{C1}=0, V_{C2}=V_\pi$); and NOR gate ($V_{C1}=V_\pi, V_{C2}=V_\pi$). In combination with the multifunctional Buffer/Inverter, the described logic gate is sufficient to construct any Boolean logic function. The ability to realize reconfigurable Majority gates gives us a powerful tool for different architecture solutions to be discussed in Section IV.

III Experimental Data and Numerical Simulations

In order to illustrate the performance of the basic elements and comprehend the magnitude of the inductive voltage produced by the spin wave signal, we present experimental data on spin wave excitation, transport, modulation, and detection in 100nm thick CoFe film. The general view of the device structure is shown in Fig.3(a). A $Co_{30}Fe_{70}$ film (saturation magnetization of 2.2 T) was deposited by high vacuum rf-sputtering system on silicon substrate and covered by 300nm silicon oxide. There are two microstrips (asymmetric coplanar strip transmission lines), on the top of the structure. One microstrip is the transducer to excite spin waves, and the other one is the receiver to detect the inductive voltage produced by the propagating spin waves in the film. The complete experimental setup is shown in Fig.3.(b) It consists of Picosecond 4000B Pulse Generator to bias the excitation antenna, device structure, and 50 GHz oscilloscope connected to the receiving microstrip. The setup design is typical for the time-resolved inductive voltage measurement technique which has been successfully applied for spin wave propagation study in nanometer thick ferromagnetic films [24, 25]. The distance between the microstrips is 4μm. The input signal (to the transducer) rising time is 1.2 ns, and the maximum input amplitude is 10.0 V. In Fig.4(a) we plotted the inductive voltage detected in the receiver microstrip as a function of time obtained for 50 Oe external magnetic field applied in the direction perpendicular to spin wave propagation. To exclude the effect of the direct inductive coupling between the antennas, we carried out the experiments with and without magnetic field and made a subtraction for each curve as described in [25]. All data are obtained at room temperature. The output signals have form of damping harmonic oscillations. Both amplitude and phase of the output signal depend on the strength of the external magnetic field. In Fig.4(b) we plotted the amplitude and the phase of the output signal obtained at different values of the bias magnetic field. From these data, one can estimate several important parameters: signal propagation speed $v_g \sim 10^4$m/s, amplitude attenuation time $\tau \sim 0.8$ns, signal modulation efficiency $\Delta\varphi/\delta H \sim 30^0/15$Oe. These numbers are typical for spin wave transport in conducting ferromagnetic nanometer thick films at low bias magnetic fields at room temperature [24-26], and can be used as the reference points for further consideration. The coupling efficiency $V_{out}/V_{in} \sim 0.001$ is pretty low for the studied device and can be enhance by optimizing the microstrip configuration. It should be noted, that these parameters are defined by different physical mechanisms and are subject for significant variation for different material structures and propagation conditions. Signal propagation speed is defined by the waveguide material, spin wave frequency, magnitude and direction of the external magnetic filed. For example, spin waves propagating along the direction of the static magnetic field (backward volume magnetostatic spin waves) or perpendicular to the field (magnetostatic surface spin waves) have significantly different (about 20 times) group velocities [26]. Attenuation time and signal modulation efficiency depends both on the thickness of the magnetic waveguide, signal frequency and external magnetic field. The coupling efficiency is mainly defined by the design of the spin wave generating/detecting contours. The key physical property, which makes spin wave utilization feasible, is the long spin wave *coherence length*. As it was experimentally demonstrated in spin wave Mach–Zehnder-type interferometer, spin wave signals remain coherent after propagation of millimeters distance in yttrium iron garnet film [18]. In the permalloy films[25], the coherence length is of the order of tens of microns at *room temperature*.

There is a few experimental works demonstrating magnetization easy axis rotation in the two-phase piezoelectric/ferromagnetic structures [32]. In order to illustrate the performance of the magnetoelectric cell, we present results of numerical simulations for the structure shown in Fig.5. There is ferromagnetic film polarized in the *x-y* plane with a region covered by piezoelectric material. Hereafter, we will refer to this region as a magnetoelectric cell and specially define the magnetization in this region as the "cell magnetization". The propagation of spin waves in a ferromagnetic layer is modeled using the Landau-Lifshitz-Gilbert equation:

$$\frac{d\vec{m}}{dt} = -\frac{\gamma}{1+\alpha^2} \vec{m} \times \left[ \vec{H}_{eff} + \alpha \vec{m} \times \vec{H}_{eff} \right], \tag{1}$$

where $\vec{m} = \vec{M}/M_s$ is the unit magnetization vector, $M_s$ is the saturation magnetization, $\gamma$ is the gyro-magnetic ratio, and $\alpha$ is the phenomenological Gilbert damping coefficient. The first term of equation (1) describes the precession of magnetization about the effective field and the second term describes its dissipation. $\vec{H}_{eff}$ is the effective field given as follows:

$$\vec{H}_{eff} = \vec{H}_d + \vec{H}_{ex} + \vec{H}_a + \vec{H}_b, \tag{2}$$

where $\vec{H}_d$ is the magnetostatic field ($\vec{H}_d = -\nabla\Phi$, $\nabla^2\Phi = 4\pi M_s \nabla \cdot \vec{m}$), $\vec{H}_{ex}$ is the exchange field ($\vec{H}_{ex} = (2A/M_s)\nabla^2 \vec{m}$, $A$ is the exchange constant), $\vec{H}_a$ is the anisotropy filed ($\vec{H}_a = (2K/M_s)(\vec{m} \cdot \vec{c})\vec{c}$, $K$ is the uniaxial anisotropy constant, and $\vec{c}$ is the unit vector along with the uniaxial direction), $\vec{H}_b$ is the external bias magnetic field. We assume that the applying voltage across the region with piezoelectric material induces stress, and, thus, locally rotates the easy axis in the *x-y* plane as shown in Fig.5. We assume that $\vec{c} = (0, \vec{y}, 0)$ for cell at $V_G=0$, and $\vec{c} = (\vec{x}, 0, 0)$ at $V_G = V_A$. If there is no gate voltage applied, all domains in the ferromagnetic film are aligned along with the easy axis and piezoelectric material has no effect on spin wave propagation in the ferromagnetic film. In Fig.6 we present the results of numerical simulations showing spin wave signal propagation in the ferromagnetic film at $V_G=0$. For simplicity, we introduce a small perturbation of the domain polarization $M_x/M_s=0.1$ at the left edge of the film at initial moment of time, while all other domains in the film are completely polarized along with the *y* direction $M_y/M_s=1$. The relaxation of the edge domain produces a spin wave packet, which propagates from the left to the right through the film. In our numerical simulations we used $\gamma=2.0\times10^7$rad/s/Oe, $4\pi M_s=10$kG, $2K/M_s=4$Oe, $H_{ext}=5$Oe, and $\alpha=0.0097$ which are typical parameters for permalloy films [24, 33]. In Fig.6 we depicted magnetization change in time at the distance 5µ from the edge. The amplitude of the spin wave signal decreases with the decay time $\tau$ given by $\tau = (2\pi\gamma\alpha M_s)^{-1}$. After a sufficiently long time, all domains in the film become polarized along with the *y*-direction ($M_x/M_s=0$ everywhere in the film). Next, we consider the situation when the gate voltage is applied, and the easy axis orientation is locally changed from *y* to *x* direction under the piezoelectric material (magnetoelectric cell). In this case, the cell magnetization has two preferable directions: along with or opposite to the *x* axis. The aim of our simulations is to show, that the magnetization direction of the cell can be controlled by the *phase* of the incoming spin wave signal. Without loss of generality and in order to emphasize the effect of easy axis rotation, we assume that the uniaxial anisotropy constant *K* significantly increases under the applied voltage, so that $2K(V_A)/M_s$=40 Oe, and the anisotropy filed $H_a$ is much higher than the external bias field $H_b$, $H_a>>H_b$. In Fig.7 we show the final (at the time much longer than required for spin relaxation) magnetization distribution ($M_x$) along the ferromagnetic film. We considered two cases when the incoming spin wave signal has two initial phases 0 and π, respectively. It is shown, that the final cell magnetization is along with *x* axis if the initial phase is 0, and opposite to the *x* direction if the initial phase is π. The inserts of the Fig.7 show the dynamics of the cell magnetization relaxation. These results illustrate the possibility of a bi-stable magnetoelectric cell switching by spin wave signal. The cell will preserve its magnetization as long as the gate voltage is applied. As the gate voltage $V_G$ is turned off, the easy axis returns to its initial orientation and the domains become polarized along with a *y* axis. The change of the cell magnetization produces a spin wave signal. In Fig.8 we have shown the results of numerical simulations illustrating cell magnetization as a function of time. Here, we start with magnetization distribution shown in Fig.7. As the gate voltage is turned off at the moment *t=0.5,* the domains under the piezoelectric film start to relax along with the *y* direction producing a spin wave signal as it is shown in Fig.8.

Based on the presented results of numerical modeling, we want to outline several key properties of the magnetoelectric cell operation: (i) the magnetoelectric cell can serve as a spin memory as long as the gate voltage is applied; (ii) it can be used as a spin wave amplifier as the amplitude of the output spin wave signal may be manifolds higher than the amplitude of the input spin wave signal; (iii) it is possible to equalize the amplitude of the spin wave signals using cell magnetization saturation (the magnetization is limited by $M_s$). Potentially, it is also possible the use magnetoelectric cells for spin wave excitation and modulation. All these makes the magnetoelectric cell a promising multifunctional element to be exploited in a variety of architecture solutions.

IV. Possible Architectures

The building blocks described in Section II may be arranged in many possible ways leading to a plethora of architecture solutions some of which has been described in Ref.[10]. In this section we will emphasize the potential advantages of using spin waves for information transmission and processing at the architecture level and illustrate it on three specific examples.

The wave nature of the magnetic wave let us excite, transmit and process a number of signals in one structure at the same moment of time. For each frequency, a bit of information can be encoded into a phase as described in Sec.II. Because of the spin wave dispersion, spin waves of different frequencies have different group velocity and can be modulated and processed independently by using a space distributed set of modulators and magnetoelectric cells. In Fig.9 we show a multi-bit processor scheme comprising of the converters, a set of modulators and magnetoelectric cells arranged in a sequence on the same spin wave bus. One micro-antenna, can be used to excite a number of spin waves of different frequencies. Then, the time and duration of the modulation voltage(current) in each modulator can be adjusted to provide a $\pi$–phase shift for any given frequency. Similarly, the second ACPS line is used to detect the output inductive voltage produced by multiple spin wave signals. Each modulator and magnetoelectric cell can be adjusted to work on any specific frequency to reduce the total number of devices in the array. We note that simultaneous propagation of a number of waves may result in a nonlinear phenomena of magnetic media, such as frequency up ($2f_1, 2f_2...2f_n$) and down ($0.5f_1, 0.5f_2...0.5f_n$) conversion and the appearance of the mixed frequency signals ($f_1+f_2$, $f_1-f_2,.. f_1+f_n$), which can be both a technological drawback and an advantage for information processing at the same time. The experimental data on frequency conversion in 100nm ferromagnetic film can be found in Ref.[34].

In principle, the ability of multi-bit processing is a significant advantage over the CMOS-based architecture where each complimentary pair can process only one bit per time. The weak point of the above described processor is in the possible inductive coupling among the modulators and input/output antennas, which can be substantial taking into account (sub)micron distance between the devices. The performance of each modulator in the array will be affected by the magnetic field produced by the other modulators. It would be of great benefit to eliminate the use of electric current for spin wave modulation and utilize only voltage driven devices similar to the proposed magnetoelectric cell in order to reduce the undesirable inductive coupling.

The next example is a Cellular Nonlinear Network (CNN) consists of a two-dimensional array of magnetoelectric cells connected with a spin wave bus. The basic idea is to make a crossbar structure consists of piezoelectric (columns) and conducting wires (rows) on the top of the ferromagnetic film as shown in Fig.10 The area of intersection between the wires automatically forms a magnetoelectric cell. Each metallic wires on the top of the structure serves for bias a large number of cells. Input/output data is transmitted via the converters placed on the edge of the structure. By applying bias voltage to the top contacts, we can activate (change easy axis) of the magnetoelectric cell in the array. The polarization of the cell (along or opposite to the new easy axis) is defined by the sum of the spin waves coming from the neighbor cells. Thus, each cell acts as a majority gate with multiple inputs. The radius of interaction among the cells in the network is limited by the spin wave attenuation length in the given ferromagnetic material (at least tens of microns at room temperature). The principle of operation of the CNN is similar to one described in Ref [35]. Input data can be read-in in a row-by-row sequence manner. Then, different transition rules can be applied by manipulated bias voltages. Final read-out operation is accomplished in a sequence similar to the read-in.

This spin-wave based CNN preserves all advantages of the magnetic Cellular Automata [8, 36] and provides some additional advantages. From the fabrication point of view, it is convenient to make a crossbar structure *to create* (and bias!) a two-dimensional cell array. The defect tolerance may be enhanced manifolds in comparison to the domain-based structure, as the interaction among the cells is via spin waves and any defect with characteristic size much smaller than the wavelength is permissible. The utilization of the common ferromagnetic film to connect logic cells in the 2-D array provides another significant advantage, as the interaction among the cells is not limited to the nearest neighbors only. As the phase of the spin wave signal is a function of the bias magnetic field, the phase of the spin wave signals coming from the distant cells can be controlled by the bias magnetic field. It provides an intrigue possibility to control logic functionality by a global parameter – bias magnetic field applied to the whole CNN structure. For example, for the fixed distance among the nearest neighbor cells, the phase of the spin wave signals may be 0 or $\pi$ depends on the bias magnetic field. The latter translates in the possibility of realization two useful image processing functions: noise reduction and edge detection, where the cell changes its state to the majority or opposite to the majority of states of its neighbors [37]. The spin-wave based CNN has many potential advantages already described in Ref. [38], and requires further investigation.

Field Programmable Gate Array (FPGA) is feasible by combining reconfigurable majority logic gates described in Section II. In Fig.11(a) we have shown a material of a three-input three-output reconfigurable majority gate which consists the waveguide structures comprising splitters and combiners and a magnetoelectric memory cell. Each ferromagnetic waveguide line is combined with a metallic wire separated by an insulator layer. The wires are aimed to serve as individual modulators for each ferromagnetic line.

The number of inputs and outputs for the majority gate can be varied. For example, in Fig.11(b) we show a FPGA block consists of thee reconfigurable majority gates (three, five, and seven inputs, respectively) depicted as (MAJ*) connected in series. By using modulators, each block can be programmed after the manufacturing to perform custom designed logic functions. The FPGA scheme can be further sophisticated in many ways by including modulator in the inter-block interconnect waveguides and increasing number of input/output ports and/or increasing the number of reconfigurable gates in each block. The key advantage of the proposed FPGA architecture is in the Majority logics. Majority gate is an important element in fault-tolerant computing and other applications [39]. CMOS-based technology has significant difficulty in the implementing of majority gates, as the number of transistors per majority gate scales proportional to the number of inputs [40]. In contrast, a wave-based approach (in particularly spin wave-based one) is free of this constrain. A number of waveguides can be combined with one magnetoelectric cell leading to a majority gate function as illustrated in Section III. Moreover, by using a three-terminal majority gate with one control input, it is possible to realize AND and OR gates in the same device structure. In combination with NOT gate (described in Section II), any Boolean function can be built.

V. Discussion

The proposed magnetic NanoFabric is based on spin-wave devices and possesses its advantages and shortcomings. On one hand, the utilization of the spin waves of submicron wavelength provides an intrigue opportunity to realize a wave-like computer (similar to the optical computer) at the nanometer scale. Spin waves can be efficiently directed by magnetic waveguides and modulated by the applied magnetic field or by electric field via the magnetoelectric effect. There is a mechanism for direct conversion from voltage-to-spin waves and vice versa, which makes spin-wave based circuits compatible with conventional electron-based devices. On the other hand, there are some fundamental drawbacks inherent to spin waves, which will limit the performance of the spin wave-based devices. These disadvantages are (i) low group velocity, and (ii) short decay time for propagating spin wave at room temperature. Spin wave dispersion depends on the waveguide geometry, strength of the bias magnetic field, and varies for spin wave modes. In the best scenario, spin wave signal is three orders of magnitude slower and than the photons in silica or electromagnetic wave in a copper coaxial cable. The use of spin waves for information transmission implies a signal delay, which is $l/v_g$, where $l$ is the propagation distance. The disadvantage associated with low group velocity is partially compensated by short (submicrons) propagation distances resulting in 0.1-1.0ns time delay per each logic gate.

Another important disadvantage is associated with the spin wave signal damping during the propagation in the spin wave bus. The damping is caused by magnon-magnon, magnon-phonon scattering as well as the effect of the Eddy currents in conducting magnetic materials. For example, spin wave damping time in 100nm thick NiFe film is about 0.8ns at room temperature [25]. It means that a significant portion of the spin wave energy will be dissipated in the waveguide structure. It should be noted that spin wave buses are inferior to ordinary copper wires in all figures of merit and can not be considered as an alternative to meal conductors for electric signal transmission [41].

An important question to ask is whether or not spin-wave based logic circuit can be less power dissipating than performing the same function CMOS-based circuit? The total power consumption of the magnetic circuit can be expressed as follows:

$$E = N_{bit} \cdot E_C + N_M \cdot E_M + N_H \cdot E_H, \tag{3}$$

where $N_{bit}$ is the number of bits, $E_C$ is the energy per bit conversion ($E_C = E_{sw}/\eta$, $E_{sw}$ is the spin wave energy, $\eta$ is the conversion efficiency), $N_M$ is the number of modulators is the circuit, $E_M$ is the energy per one bit modulation ($E_M = I_M^2 \cdot Z \cdot t_M$, $I_m$ is the modulating electric current, $Z$ is the modulator impedance, $t_m$ is the modulation time), $N_H$ is the number of magnetoelectric cells per circuit, and $E_H$ is the energy per cell per one operation ($E_H = 0.5 \cdot C_H \cdot V_A^2$, $C_H$ is the magnetoelectric cell capacitance). The consumed energy is dissipated in the waveguide structure, the conducting wires used for magnetic field generation, and in the magnetoelectric cells (in both piezoelectric and ferromagnetic materials). There is a radical difference in the operation principles of the conventional CMOS-based circuit and a spin-wave based one. In conventional CMOS circuits, where the output voltage from one transistor modulates the conductance of the following ones, the amplitude of the voltage signal transmitted via the wires is defined by the MOSFET threshold voltage. In contrast, the energy of the spin wave signal $E_{sw}$ is not directly related to the voltage required for signal modulation. The energy of the spin wave signal is limited by the thermal noise only and can be any close to *kT*. For acceptable error rate, it can be estimated as *20kT* per bit. Energy losses during signal conversion and modulation depends on the waveguide material structure, magnetic field generating device structure and modulation time, and can be roughly estimated of the order of *100kT*. Detailed numerical assets on power dissipation in ferromagnetic spin wave bus during signal propagation and phase modulation are given in [41]. For the magnetoelectric cell, we estimate the energy required for easy axis 90 degree rotation $E_H$ ~ $10^6 kT$ using the following data: $V_A$=15V [23], and assuming $C_H$ = 1.6$10^{-16}$F (cell area $S$=100nm$^2$, piezoelectric layer thickness $d$=100nm, and dielectric constant $\varepsilon$=2000). Theoretically, the energy consumed by the magnetoelectric cell can be reduced by scaling down the area of the cell, and/or by using more efficient magnetic/piezoelectric pairs. The lack of experimental data on voltage-controlled easy axis rotation does not allow us to conclude on the practically achievable minimum. So far, it is not clear whether or not individual spin wave-based devices such as modulator and magnetoelectric cell can consume less power than a CMOS.

However, there is a fundamental advantage of using spin waves for information transmission and processing that makes possible to construct some logic gates with less number of devices than required for CMOS-based circuit. Majority gate is an example of efficient logic gate construction illustrating this advantage. Encoding a bit of information into the phase of the spin wave signal, let us exploit spin wave superposition for Majority gate construction as described in Section II. A large number of waveguides can be combined with *a single* magnetoelectric cell leading to the Majority gate operation. The whole gate can be scaled down to a single ferromagnetic wire with multiple microstrips. In contrast, the number of CMOSs required for Majority gate scales proportional to the number of inputs. Majority logic is a way of implementing digital operations in a manner different from that of Boolean logic. In general, Majority logic is more powerful for implementing a given digital function with a smaller number of logic gates [40]. For example, the full adder may be constructed on three majority gates and two inverters (3 magnetoelectric cells and 2 modulators). In contrast, a Boolean-based implementation requires a larger circuit with seven or eight gate elements (about 25–30 MOSFETs) [32]. The main reason Majority logic has been out of stage for decades is because its CMOS realization is inefficient. Only with the development of novel devices such as Josephson junction circuits, which is not feasible at room temperature [42], and quantum cellular automata [43], the Majority logic realization became feasible. The proposed spin wave based NanoFabric offers an intrigue opportunity to make a reconfigurable Majority gate which logic operation can be controlled by the spin wave phase modulators. In turn, the integration of reconfigurable Majority gates provides a route to building both general purpose and special task architectures such as Cellular Automata and FPGA.

We emphasize the unique feature of the proposed NanoFabric of inherently combining the advantages of analog and digital computation. Some useful functions can be done in the waveguides using spin wave interference without an introducing an additional logic element and the result of computation can be transformed in the digital form by using the magnetoelectric cells. As pointed out by T. Roska [44], there are some computational algorithms, for example those for image processing and speech recognition, that can be implemented more time efficiently using waves rather than digital signals. Spin waves of different frequencies can simultaneously excited, transmitted and modulated in the same structure, which is of great value for multi-bit parallel processing. For example, image processing function labeling can be done efficiently with $O(\log N)$ time for any given $N \times N$ image using spin wave architecture, as compared with CMOS with $O(N)$ [45].

Scalability and defect tolerance of the spin wave-based logic devices are defined by the same physical parameter - the spin wavelength $\lambda$. The length of the Buffer gate has to be an integer number of the wavelength. It restricts the minimum length of waveguide per circuit to at least one wavelength. The width and the thickness of the spin waveguides can be scaled down to several nanometers. The size of the magnetoelectric cell (the area covered by piezoelectric) can be as small as a one magnetic domain. The permissible size variation of the magnetoelectric cell and spin wave splitters/combiners, which does not effect logic functionality, can be estimated as $\lambda/8$. Any defect of the waveguide structure with characteristic size much smaller than the spin wavelength has no critical effect on spin wave propagation. There is a tradeoff, however, between scalability and defect tolerance. To scale down the length of the logic gate one needs to decrease the wavelength. At the same time, the shorter wavelength signal becomes more sensitive for structure imperfections. A wavelength of 100nm can be taken as a reference, while the optimum value has to be found taking into consideration particular material structure.

There are several important issues such as cross-talk among the magnetic field generating elements (converters and modulators), spin wave signal reflection and dispersion during propagation in waveguides, we did not discuss. A more detailed study of the basic elements operation including experimental verification is required. All these important problems deserve special study. In the present work, we focused our consideration on the general idea of magnetic NanoFabric as a powerful tool for reconfigurable logic circuits construction.

VI. Conclusions

We have described a magnetic NanoFabric for building reconfigurable spin-based logic circuits. The novelty of the proposed NanoFabric is in the combination of the previously developed spin wave bus approach with voltage-controlled magnetoelectric cells. The use of spin wave interference allow us to exploit the advantages of the analog computation, while the magentoelectic cells can be used as a non-linear element to digitize spin wave signals. As a result of this combination, a reconfigurable Majority logic gates can be realized with minimum basic elements. The operation of the basic elements has been illustrated by experimental data and numerical modeling. We presented examples of architectures such as Cellular Automata and Field Programmable Gate Array, which can be constructed on the base of the Nanofabric. There are two major disadvantages inherent to spin wave-based logic devices: lower propagation speed and high attenuation. Nevertheless these disadvantages, the proposed magnetic circuits may provide a substantial throughput enhancement at the same or less level of power consumption with comparison to CMOS-based circuits. The main advantage of the proposed NanoFabric as it let us construct logic gates with less number of devices than it required by using CMOSs. Potentially, magnetic NanoFabrics may find application as an interface between electron-based and spin-based logic circuits and as a complementary logic block to conventional general-purpose processors for special applications such as image processing and speech recognition.

Acknowledgments

. We would like to thank Dr. Jacob (Intel) and Dr. Ramesh (UC Berkeley) for valuable discussion, Dr. S. Wang and D.W. Lee (UC Stanford) for CoFe depostion, and Dr. J.-Y. Lee (UC Los Angeles) for device fabrication.  The work was supported in part by the Focus Center Research Program (FCRP) Center of Functional Engineered Nano Architectonics (FENA) and by the Nanoelectronics Research Initiative (NRI) - The Western Institute of Nanoelectronics (WIN).

Figure Captions

Fig.1 Basic elements with structure schematics. (a) Converter to translate input voltage pulse into spin wave signal and vice versa. (b) Spin waveguides toe transmit spin wave signal through the circuit. (c) Modulator to provide $\pi$-phase shift to the propagating spin wave. (d) Magnetoelectric cell – multifunctional element to serve as a memory and an amplifier for spin wave signal.

Fig.2 Logic gates and corresponding magnetic circuits. (a) Buffer/Inverter one-bit logic comprising of input Converter, Modulator, Magnetoelectric Cell, and output Converter. (b) Three-input Majority gate consists of Converters, Combiner, and the Magnetoelectric Cell. (c) Multifunctional logic gate built on the base of the Majority gate by introducing two Modulators. Depends on the control voltages applied to the Modulators, the gate operates as AND, OR, NAND, and NOR logic gates.

Fig.3 (a) General view of the device structure. From the bottom to the top there are silicon substrate, 100nm thick $Co_{30}Fe_{70}$ film, and 300nm silicon oxide. There are two microstrips on the top of the structure. One microstrip is the transducer to excite spin waves, and the other one is the receiver to detect the inductive voltage. The distance between the microstrips is 5μm. (b) Experimental setup for spin wave signal excitation/detection. There is a Picosecond 4000B Pulse Generator connected to the excitation antenna, device structure, and 50 GHz oscilloscope connected to the receiving microstrip.

Fig.4 (a) Experimental data: inductive voltage detected in the receiver microstrip as a function of time. The distance between the transducer and receiver is 5μm. The curve is a result of subtraction of two: with 50 Oe and without external magnetic field. The direction of the external magnetic filed is perpendicular to spin wave propagation. (b) Phase and amplitude (in the insert) of the output signal as a function of the external magnetic field. Signal frequency is 2.7GHz, all data are obtained at room temperature.

Fig.5 Schematic view of the magnetoelectric magnetoelectric cell. From the bottom to the top there are silicon substrate, ferromagnetic film, and piezoelectric material. The easy axis of the magnetization ion the ferromagnetic film under the piezoelectric is assumed to rotate as a function of the applied voltage $V_G$. The easy axis is along with the $y$ axis at $V_G=0$, and along with the $x$ axis at $V_G=V_A$.

Fig.6 Results of numerical simulations showing spin wave signal propagation through the magnetoelectric magnetoelectric cell at $V_G=0$. The distance from the point of spin wave excitation to the cell is 5μ.

Fig.7 Results of numerical simulations showing magnetization of the magnetoelectric cell at $V_G=V_A$. There are two possible direction of magnetization: along with or opposite to the $x$ axis. The direction is defined by the phase of the incoming spin wave. The insets illustrate the dynamics of switching when the control voltage $V_G$ is changing

Fig.8 Results of numerical simulations showing spin wave signal generation as the control voltage to the magnetoelectric cell is changed from $V_A$ to 0. The initial cell magnetization is shown in Fig.7.

Fig.9 (a) General view of the multi-bit processor. From the bottom to the top, there are silicon substrate, ferromagnetic film, and silicon oxide layer. On the top of the structure, there is a set of converters, modulators and magnetoelectric cells arranged in a sequence. Two ACPS transmission lines on the edges are aimed to excite and detect spin wave signals on different frequencies. The pair of modulator and magnetoelectric cell is adjusted to modulate/amplify spin wave signals on specific frequency. (b) The equivalent magnetic circuit.

Fig.10 (a) schematic view of the Cellular Nonlinear Network consists of a two-dimensional array of magnetoelectric cells united by a spin wave bus. The core of the structure consists of the silicon substrate, ferromagnetic film, and silicon oxide, similar to one for the multi-bit processor. There is a crossbar structure on the top consists of piezoelectric (columns) and conducting wires (rows). The area of intersection between the wires forms a magnetoelectric ferromagnetic/piezoelectric cell. The metallic wires on the top of the structure serves to bias the cells. Under the bias voltage applied, each cell can be in two states defined by the magnetization direction. The polarization of the cell (along or opposite to the new easy axis) is defined by the sum of the spin waves coming from the neighbor cells. Each magnetoelectric cell acts as a majority gate with multiple inputs. (b) The equivalent magnetic circuit.

Fig.11(a) The material of a three-input three-output reconfigurable majority gate. It consists of the waveguide structures comprising splitters and combiners and a magnetoelectric memory cell. Each ferromagnetic waveguide line is combined with a metallic wire separated by the silicon oxide layer. The wires are aimed to serve as individual modulators for each ferromagnetic line. (b) Schematics of the FPGA block consists of thee reconfigurable majority gates (three, five, and seven inputs, respectively) depicted as (MAJ*) connected in series. By using modulators, each block can be programmed after the manufacturing to perform custom designed logic functions.

| Basic Element /Symbol | Structure Schematics |
|---|---|
| **Converter** <br><br> **Voltage-to-Spin Wave** <br><br><br> **Spin Wave-to- Voltage** <br><br> **(a)** | 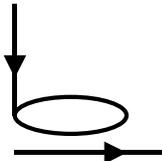 |
| **Splitter/Combiner** <br><br> **(b)** | 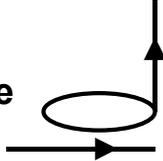 |
| **Spin Wave Modulator** <br><br> **(c)** | 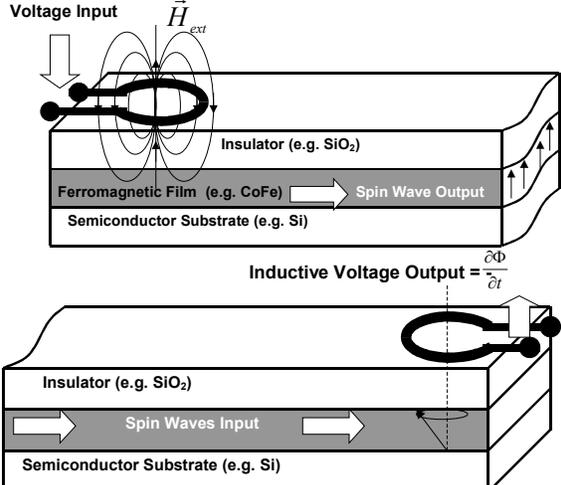 |
| **Magnetoelectric Cell** <br> (e.g. Piezoelectric-Piezomagnetic) <br><br> **(d)** | 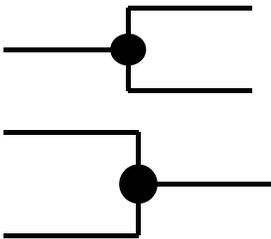 |

**Fig.1**

# Magnetic Circuits and Logic Gates

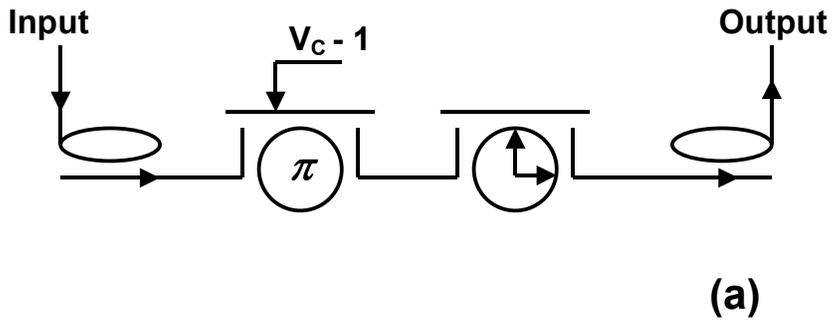

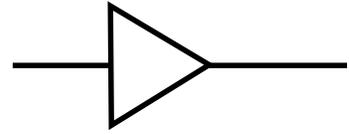

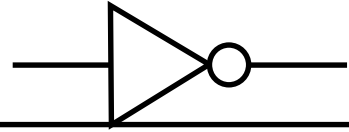

**(a)**

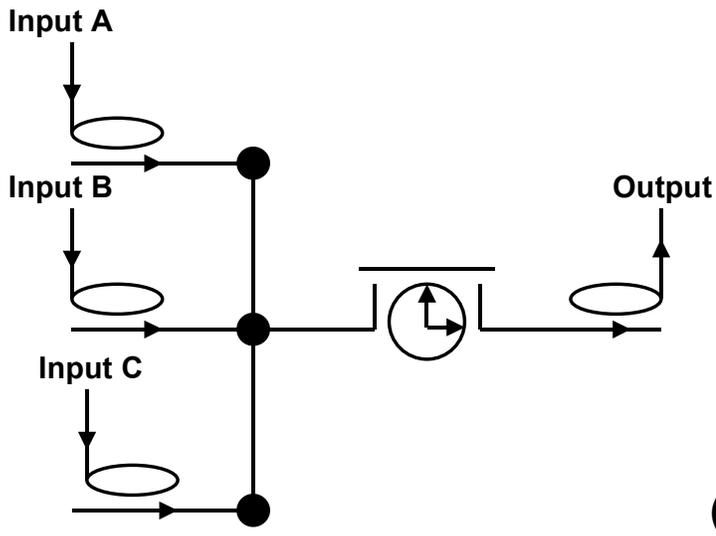

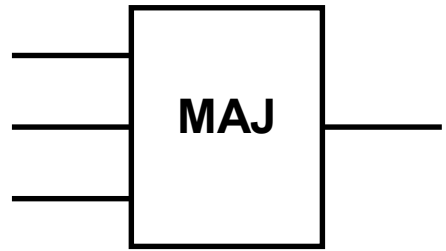

**(b)**

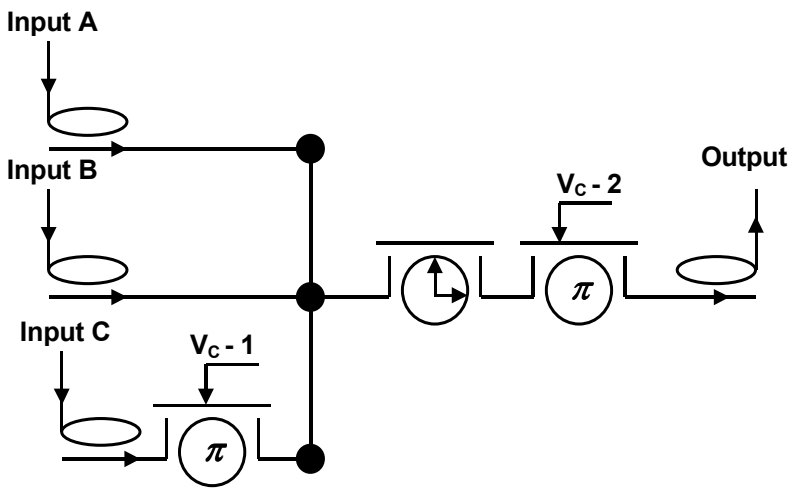

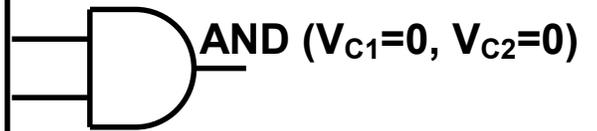

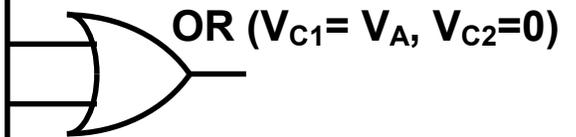

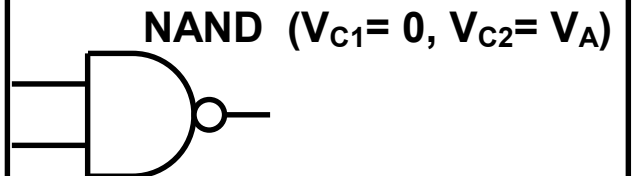

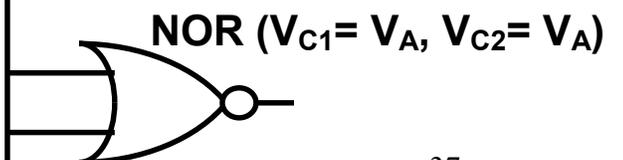

**(c)**

**Fig.2**

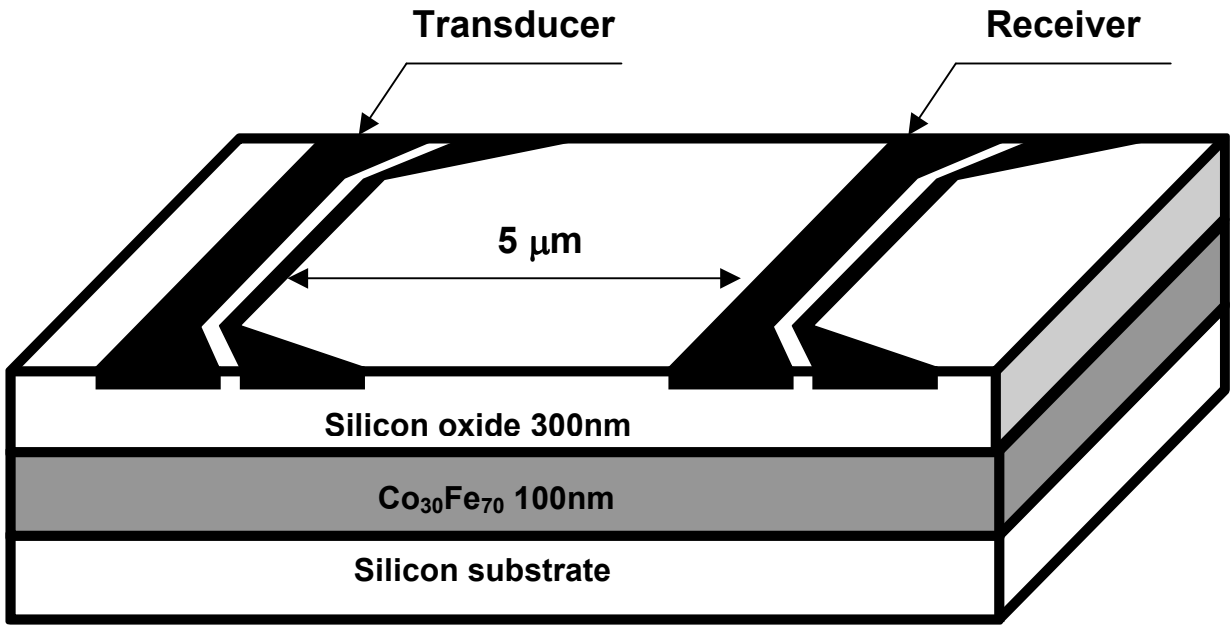
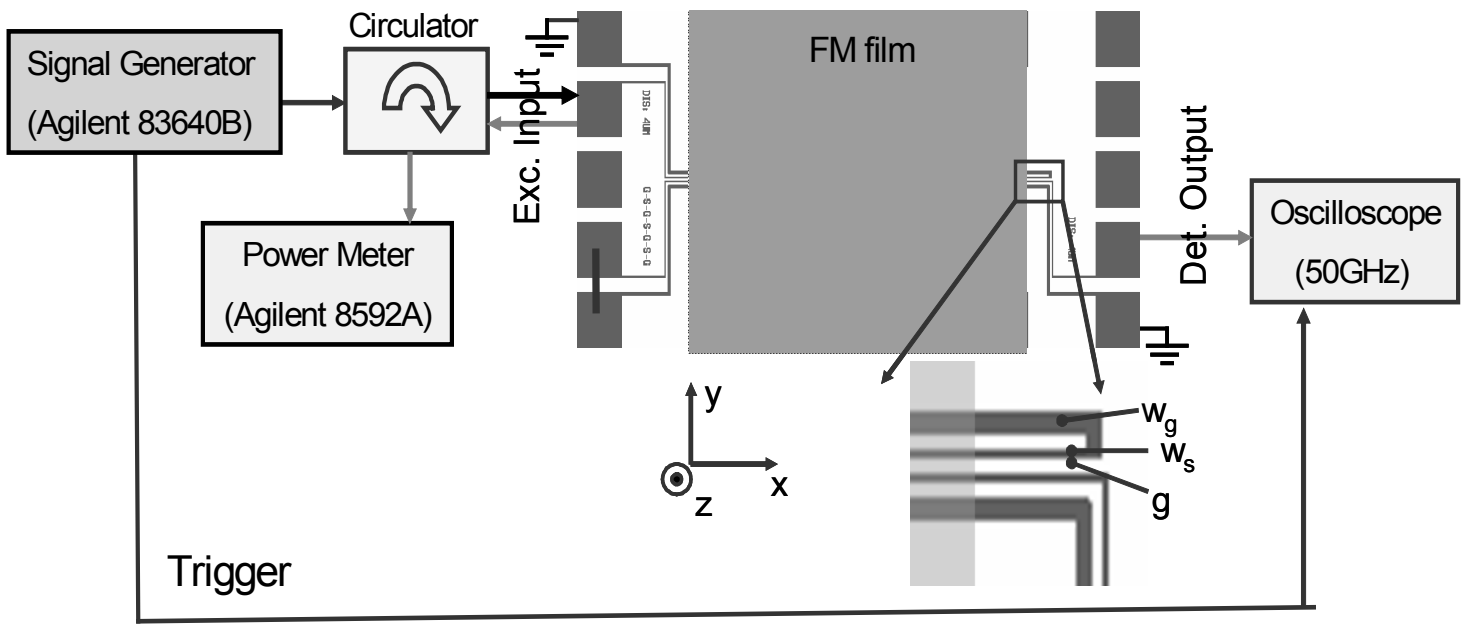

**Fig.3**

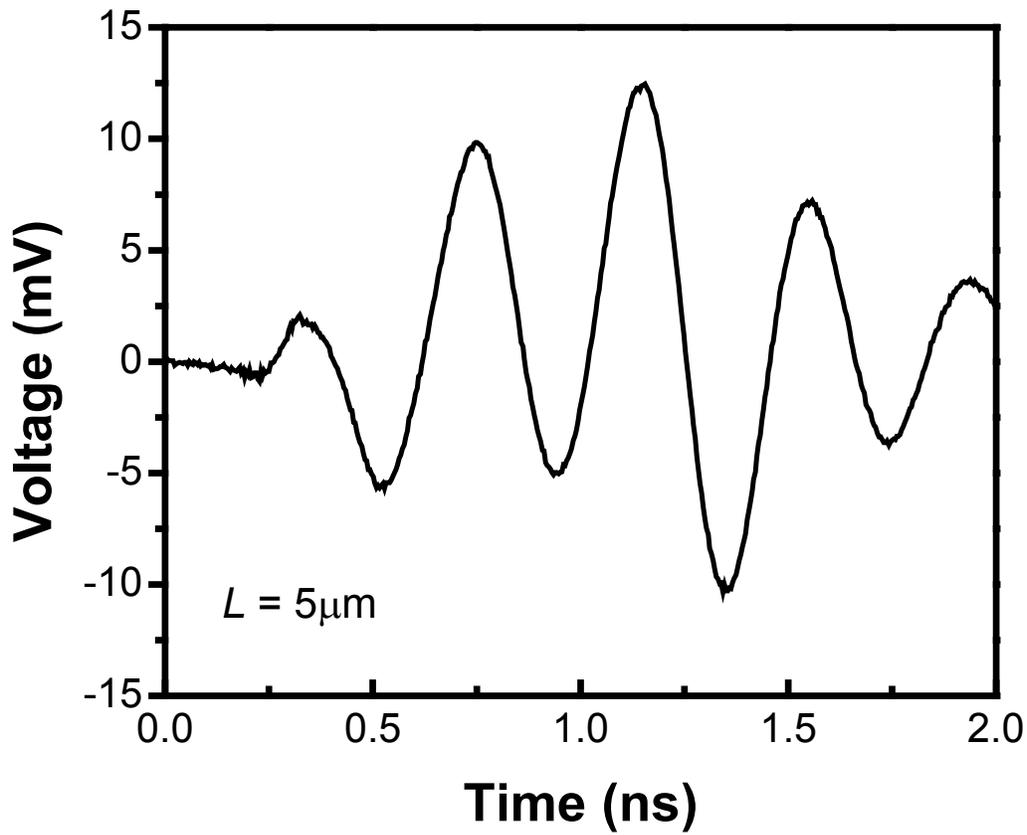

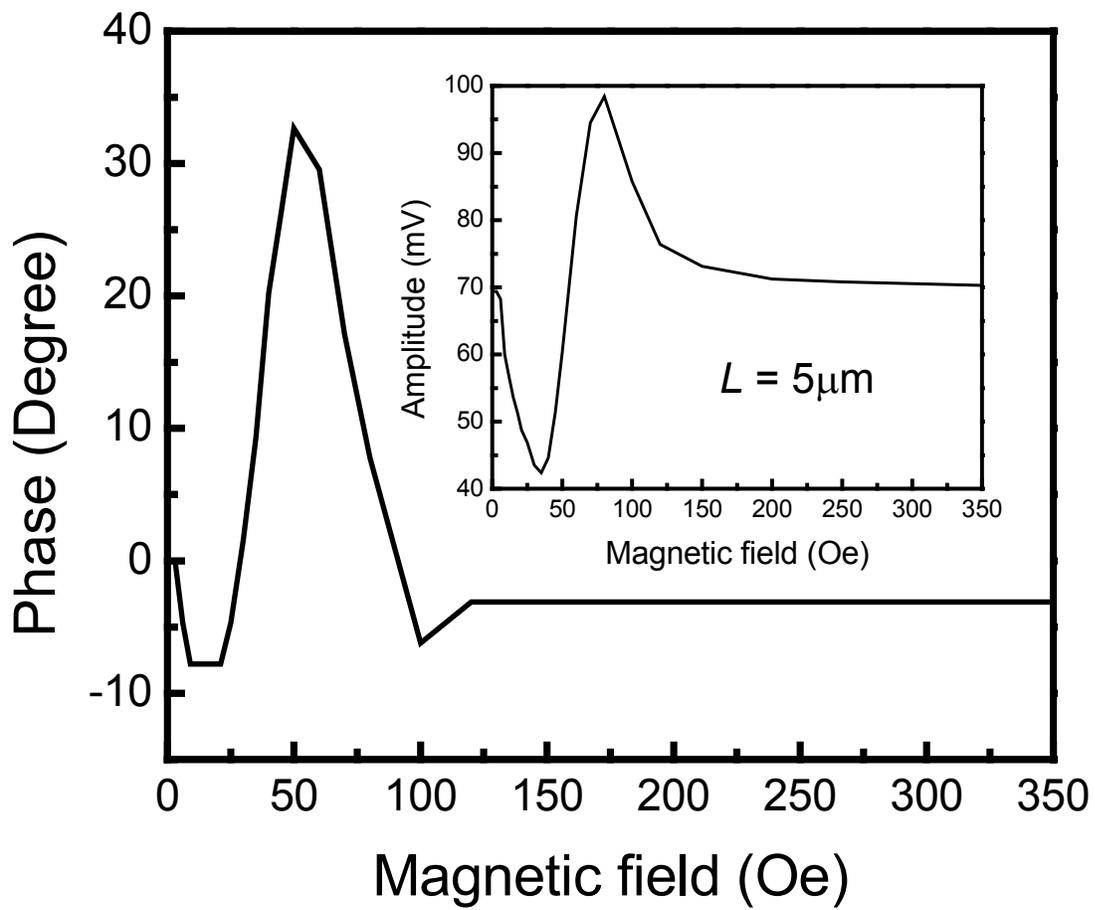

Fig.4

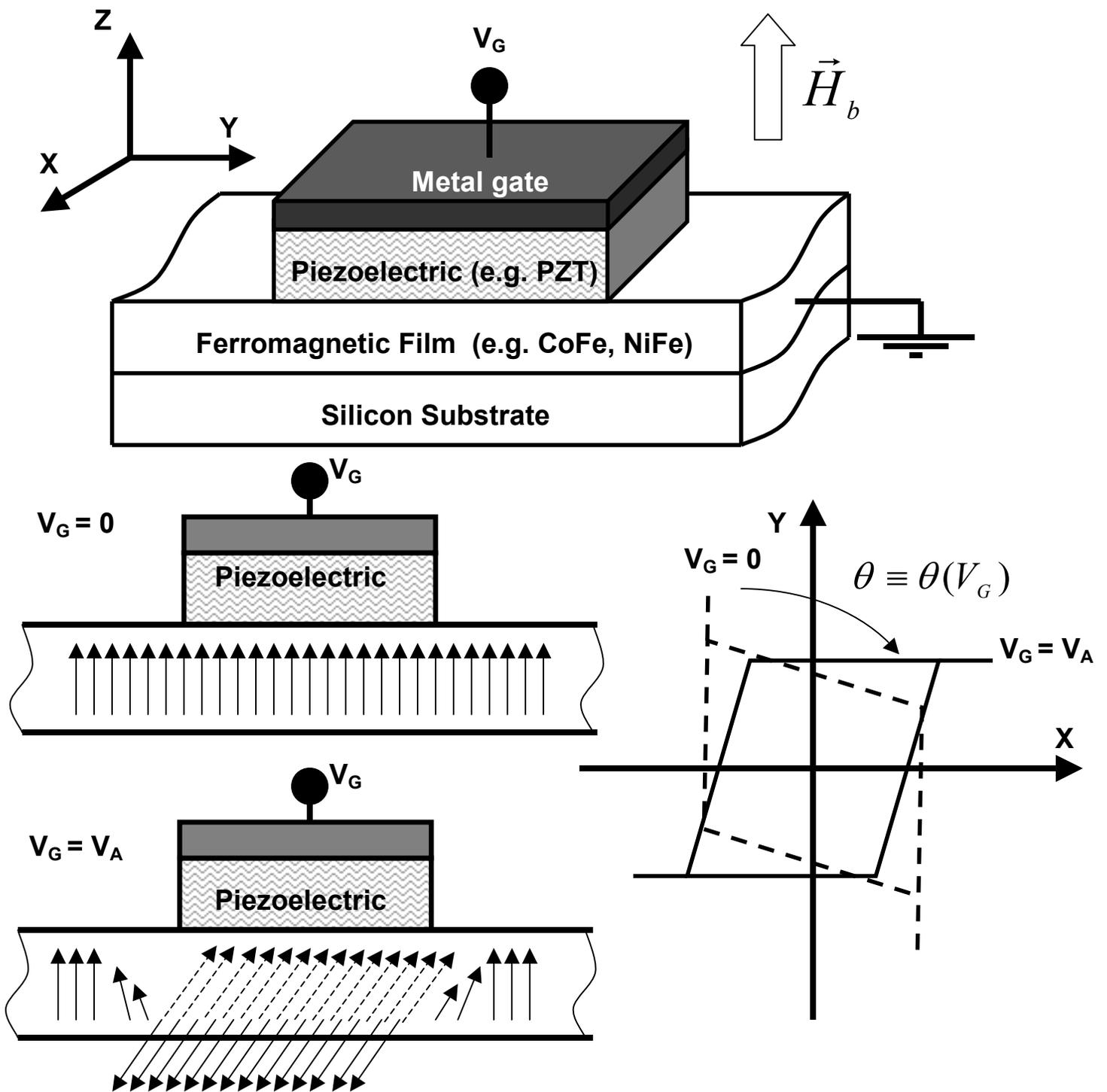

**Fig.5**

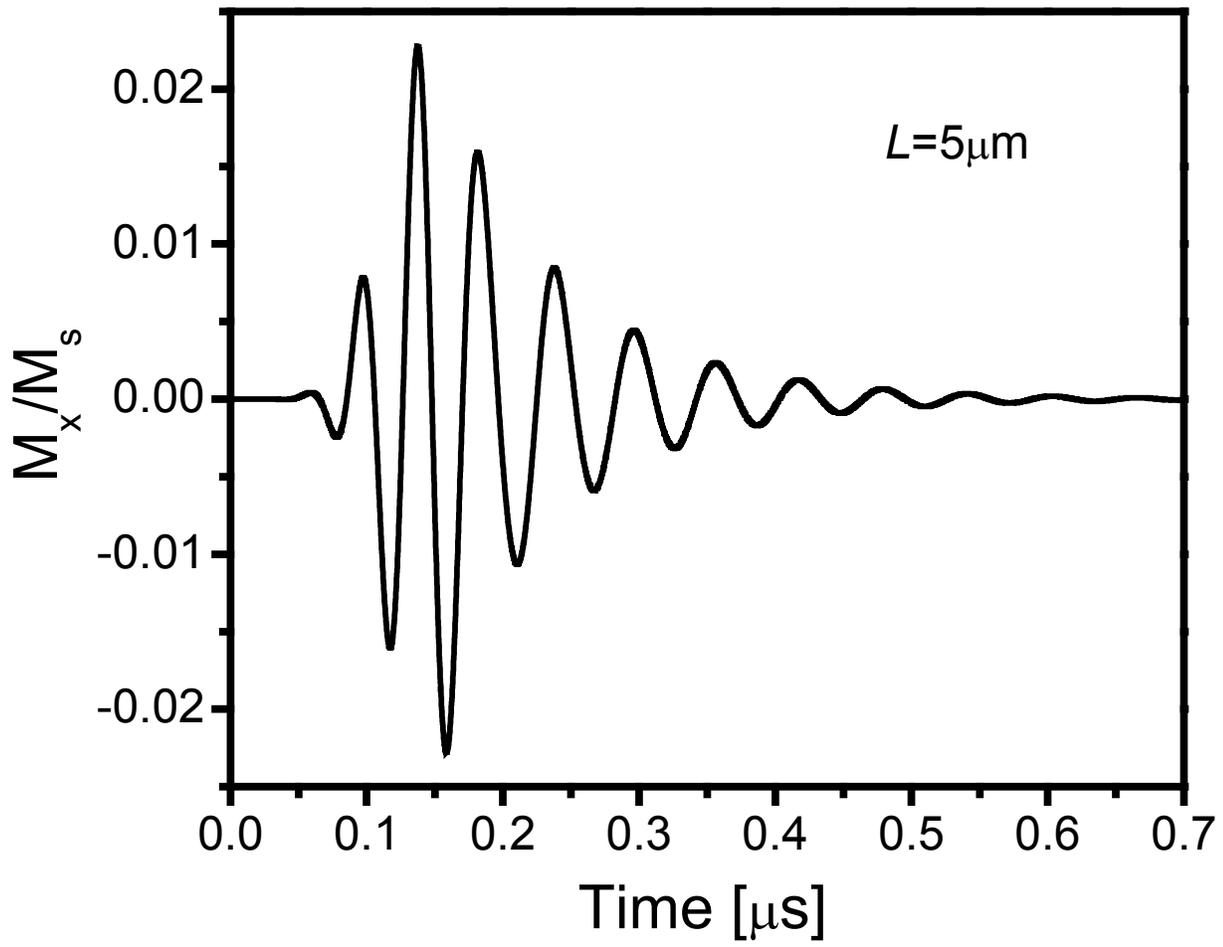

**Fig.6**

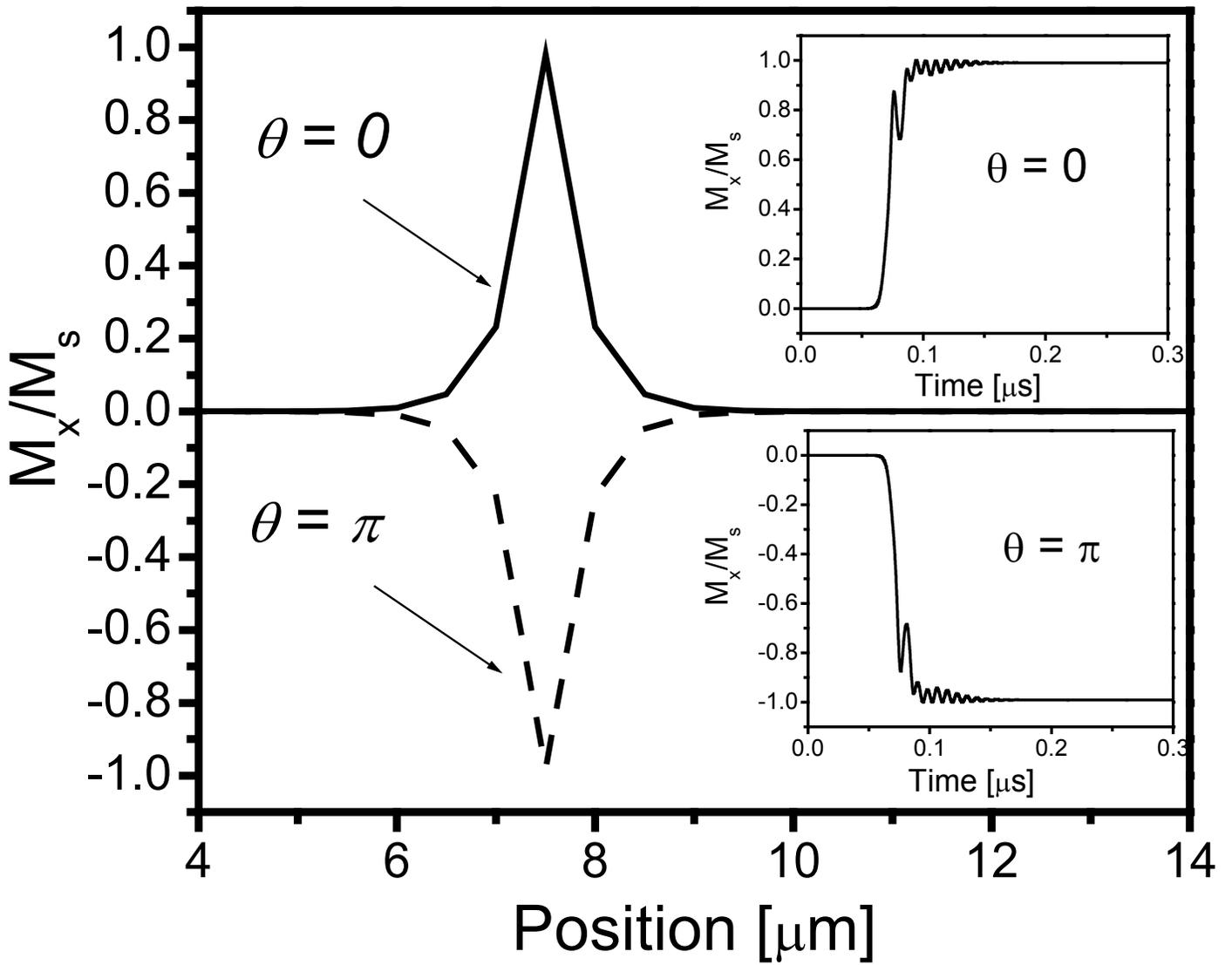

**Fig.7**

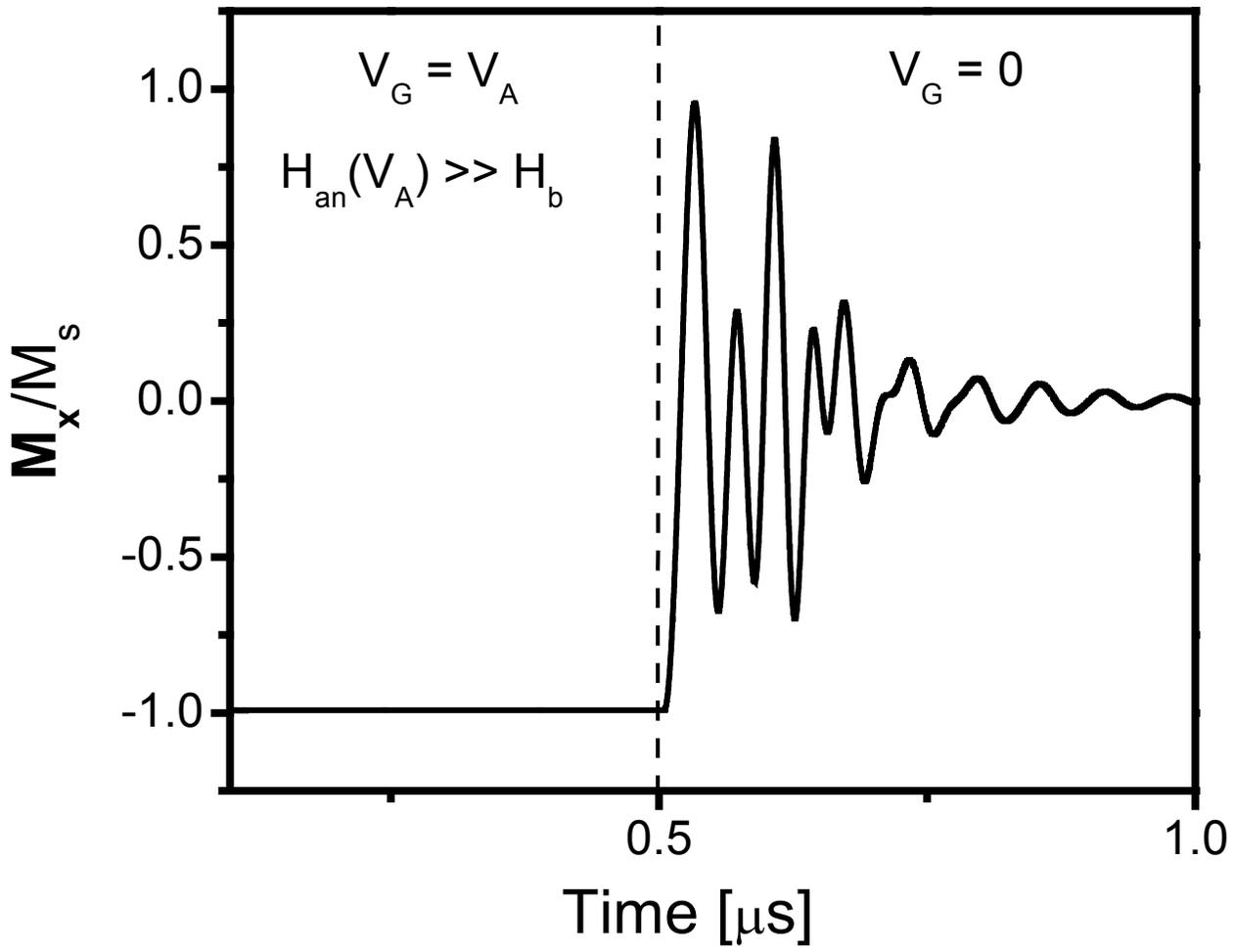

**Fig.8**

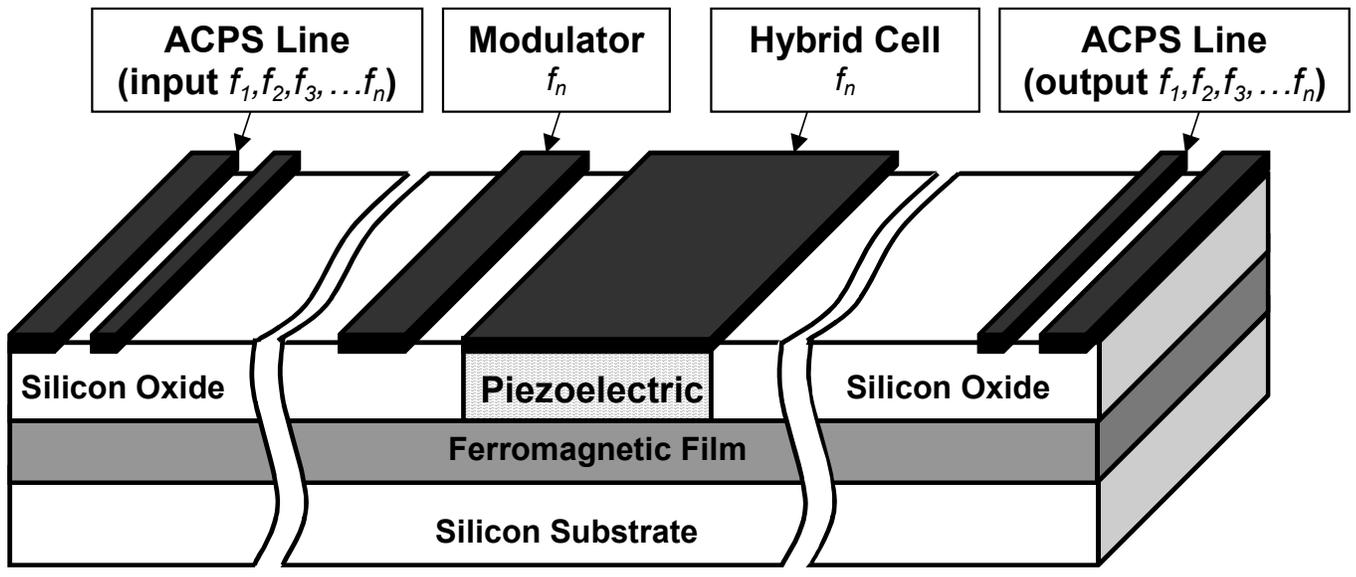

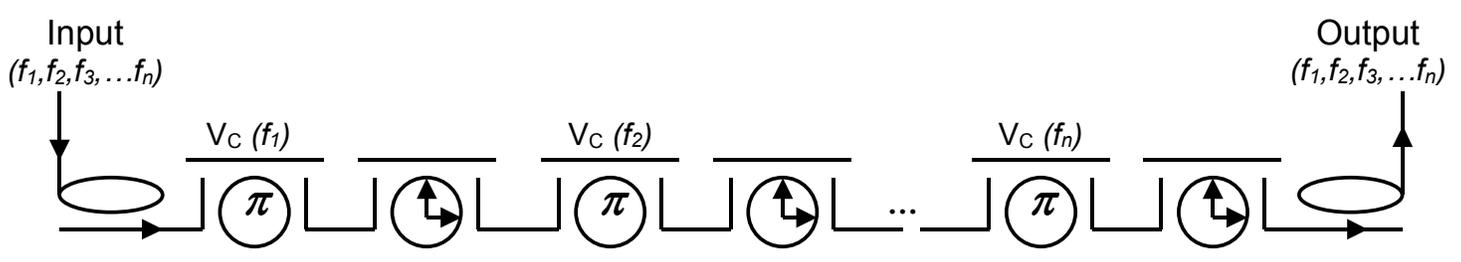

**Fig.9**

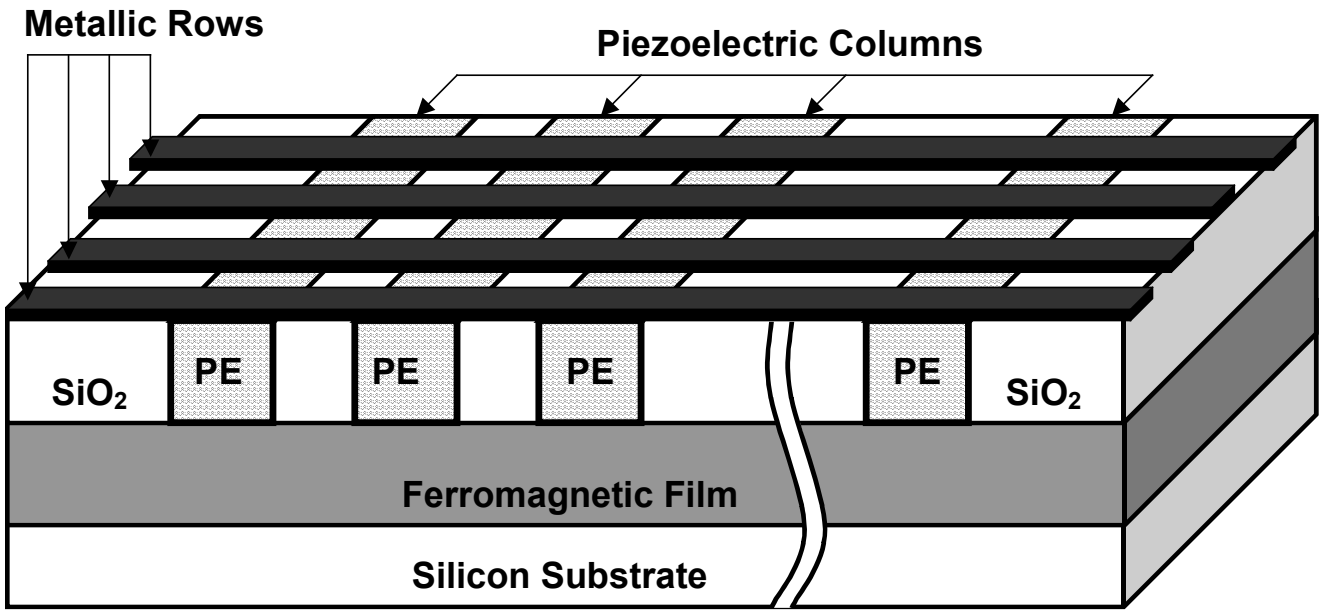
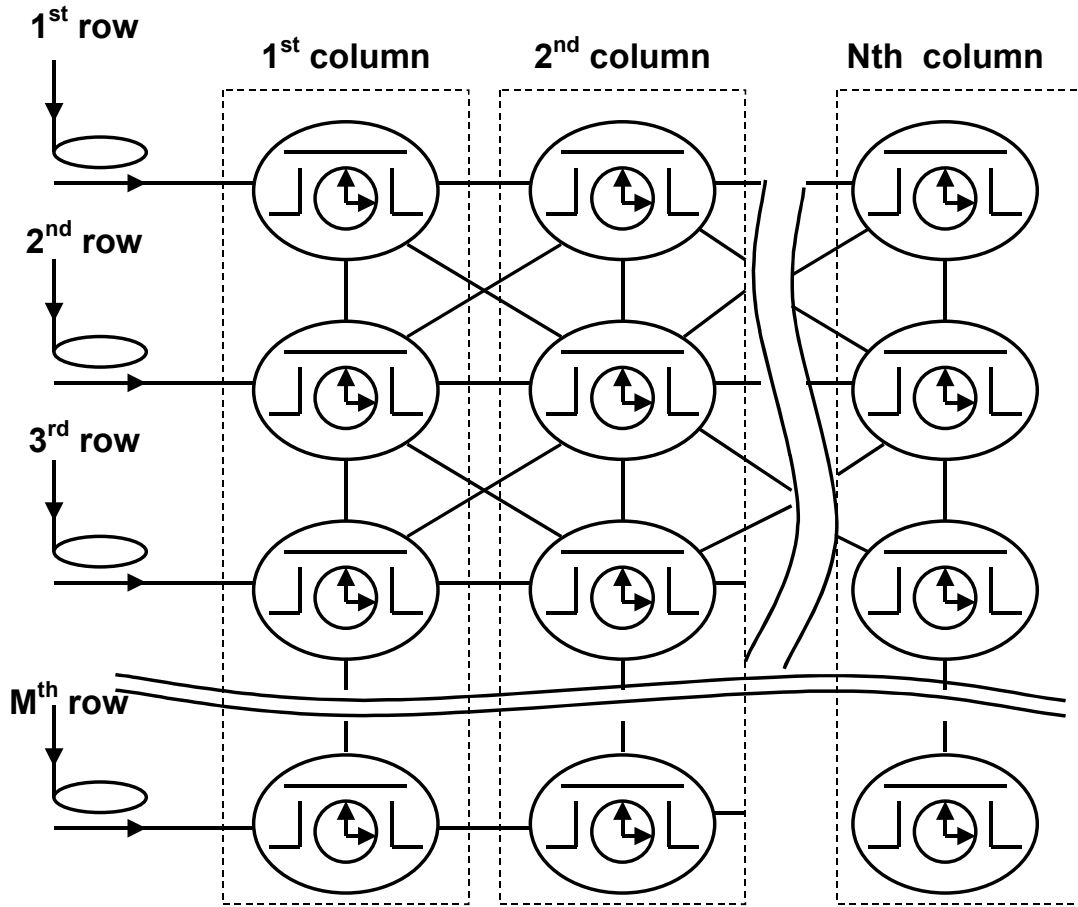

**Fig.10**         45

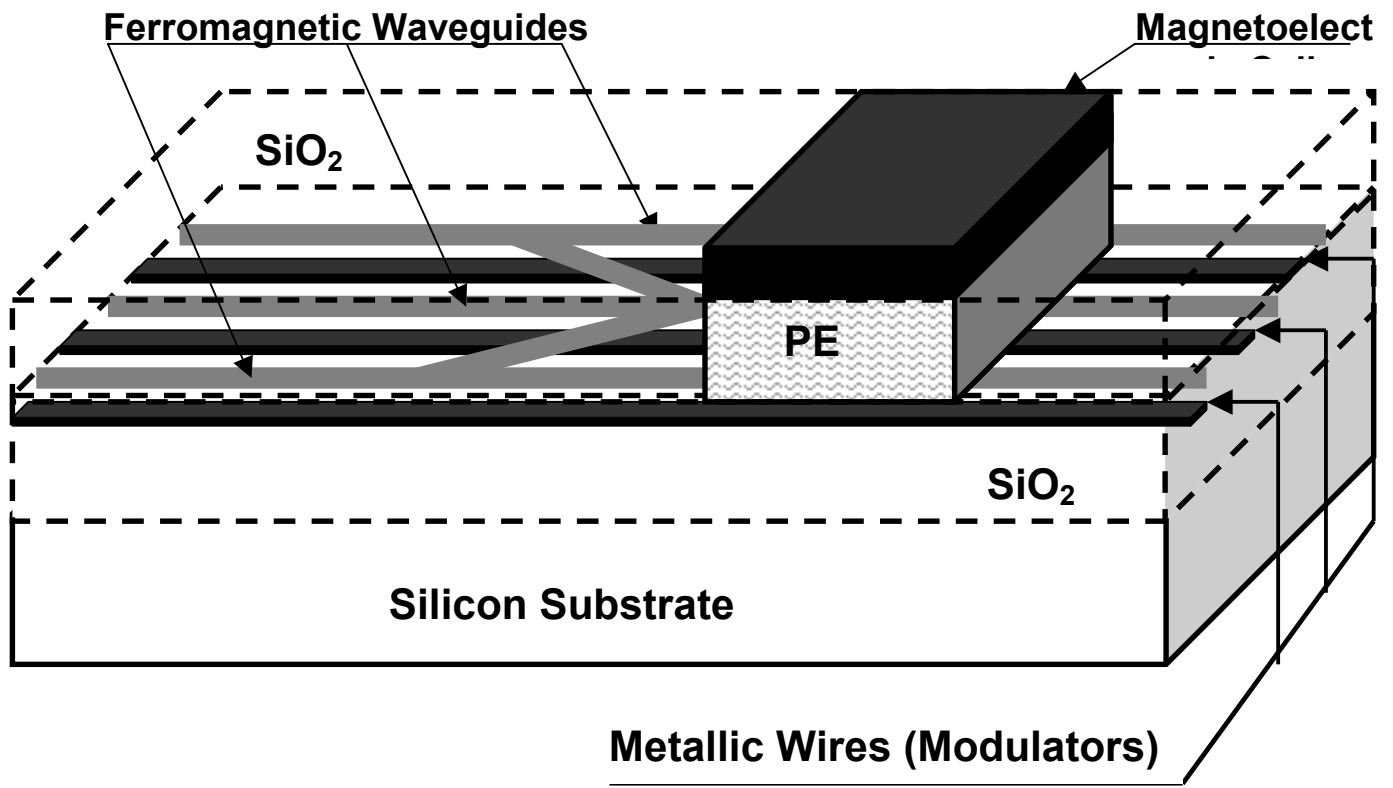

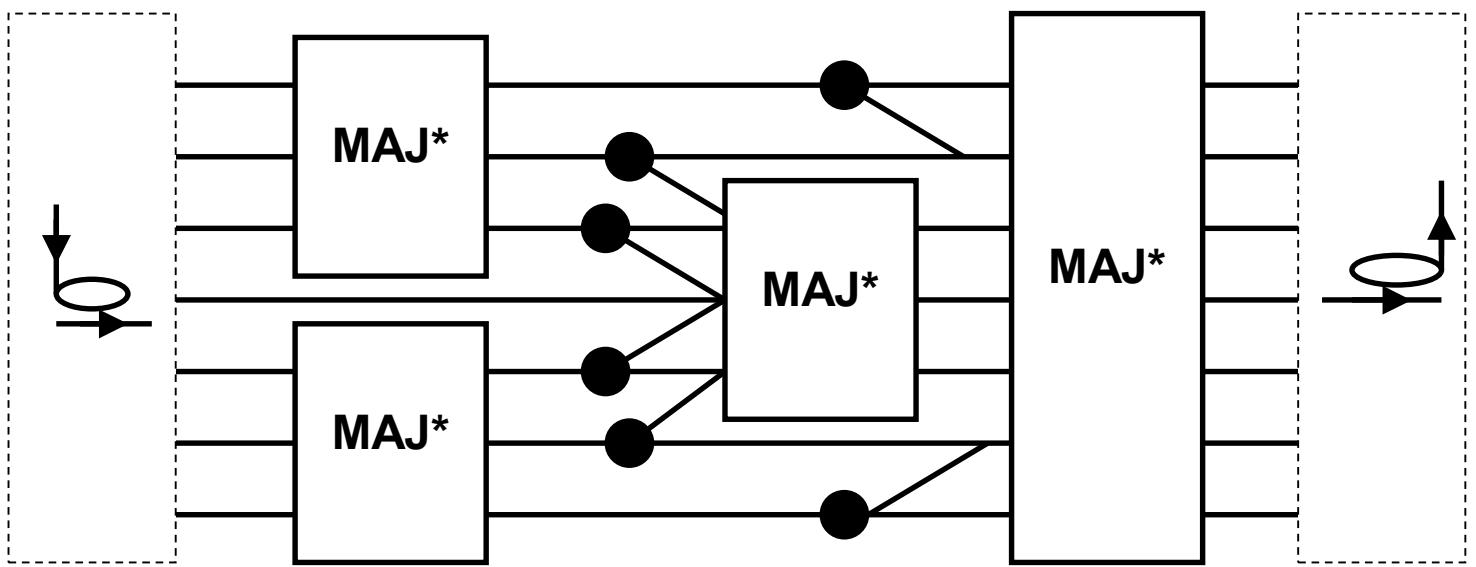

**Fig.11**